\documentclass{webofc}
\usepackage[varg]{txfonts}   % Web of Conferences font
\usepackage{wrapfig}
\begin{document}
\title{3D scaling laws and projection effects in The300-NIKA2 Sunyaev-Zeldovich Large Program Twin Samples}
%
% subtitle is optionnal
%

\author{\lastname{A. Paliwal}\inst{1}\fnsep \thanks{\email{aishwarya.paliwal@uniroma1.it}} \and
        %\firstname{E.} \lastname{Artis}\inst{2} \and
        \lastname{W. Cui}\inst{2} \and
        \lastname{D. de Andrés} \inst{2} \and
        \lastname{M. De Petris}\inst{1} \and
        \lastname{A. Ferragamo}\inst{1,4} \and
        %\firstname{G.} \lastname{Gianfagna}\inst{1} \and
        \lastname{C. Hanser} \inst{3} \and
        \lastname{J.-F. Macías-Pérez}\inst{3} \and
        \lastname{F. Mayet}\inst{3} \and
        \lastname{A. Moyer-Anin} \inst{3} \and 
        \lastname{M. Muñoz-Echeverría}\inst{3} \and
        \lastname{L. Perotto}\inst{3} \and
        \lastname{E. Rasia}\inst{5,6} \and
        %\firstname{F.} \lastname{Ruppin}\inst{8} \and
        \lastname{G. Yepes}\inst{2}
        % etc.
}

\institute{Dipartimento di Fisica, Sapienza Università di Roma, Piazzale Aldo Moro 5, 00185 Roma, Italy
\and
           Departamento de Física Teórica and CIAFF, Módulo 8, Facultad de Ciencias, Universidad Autónoma de Madrid, 28049 Madrid, Spain
\and
          Univ. Grenoble Alpes, CNRS, Grenoble INP, LPSC-IN2P3, 53, avenue des Martyrs, 38000 Grenoble, France
\and
           Instituto de Astrofísica de Canarias (IAC), C/Vía Láctea s/n, 38205 La Laguna, Tenerife, Spain
%\and
 %          High Energy Physics Division, Argonne National Laboratory, 9700 South Cass Avenue, Lemont, IL 60439, USA
\and 
          National Institute for Astrophysics, Astronomical Observatory of Trieste (INAF-OATs), via Tiepolo 11, 34131 Trieste, Italy
\and
          Institute for Fundamental Physics of the Universe (IFPU), via Beirut 2, 34014 Trieste, Italy
%\and
         %Univ. Lyon, Univ. Claude Bernard Lyon 1, CNRS/IN2P3, %IP2I Lyon, F-69622, Villeurbanne, France
}

\abstract{
The abundance of galaxy clusters with mass and redshift is a well-known cosmological probe. The cluster mass is a key parameter for studies that aim to constrain cosmological parameters using galaxy clusters, making it critical to understand and properly account for the errors in its estimates. Subsequently, it becomes important to correctly calibrate scaling relations between observables like the integrated Compton parameter and the mass of the cluster.

The NIKA2 Sunyaev-Zeldovich Large program (LPSZ) enables one to map the intracluster medium profiles in the mm--wavelength band with great details (resolution of $11 \ \mathrm{\&}\ 17^{\prime \prime}$ at $1.2 \ \mathrm{\&}\ 2 $ mm, respectively) and hence, to estimate the cluster hydrostatic mass more precisely than previous SZ observations. However, there are certain systematic effects which can only be accounted for with the use of simulations. For this purpose, we employ  {\sc The Three Hundred} simulations which have been modelled with a range of physics modules to simulate galaxy clusters. The so-called twin samples are constructed by picking synthetic clusters of galaxies with properties close to the observational targets of the LPSZ. In particular, we use the Compton parameter maps and projected total mass maps of these twin samples along 29 different lines of sight. We investigate the scatter that projection induces on the total masses. Eventually, we consider the statistical values along different lines of sight to construct a kind of 3D scaling law between the integrated Compton parameter, total mass, and overdensity of the galaxy clusters to determine the overdensity that is least impacted by the projection effect.
}
%%%%%%%%%%%%%%%%%%%%%%%%%%%%%%%%%%%%%%%%%%%%%%%%%%%%%%%%%%%%%%
\maketitle

%%%%%%%%%%%%%%%%%%%%%%%%%%%%%%%%%%%%%%%%%%%%%%%%%%%%%%%%%%%%%
\section{Introduction}
\label{intro}
Clusters of galaxies are the most massive gravitationally bound objects in the Universe, sitting at the nodes of the cosmic web \cite{cweb2}. The cluster mass function \cite{PS} is highly sensitive to the underlying cosmology. This makes the cluster mass a key parameter in the context of cluster--based cosmology. The cluster mass, however, can only be inferred through cluster observables at multiple wavelengths like X--ray luminosity, mm observation of the Sunyaev–Zeldovich (SZ) effect \cite{sz}, optical lensing effect etc., under given theoretical assumptions (see \textit{e.g.} \cite{SZmass,wl}). Another way to infer the cluster mass is by the use of scaling laws, which relate the mass to observables, typically using a power law \cite{p20,schell,aarti}. All these methods are prone to their sources of mass bias and it is critical to understand their associated astrophysical and systematic details.

The mass estimation in the above--mentioned ways is limited to a single projection available to us as observers, often forcing the assumption of spherical symmetry in order to enable de-projection to 3D quantities. This creates room for increased scatter on the mass estimates (see \textit{e.g.} \cite{miren}). Additionally, these masses are typically estimated at a fixed radius corresponding to the overdensity $\Delta=500 \ \mathrm{or} \ 200$\footnote{$\Delta = \frac{M_{\Delta}/(\frac{4}{3}\pi r_{\Delta}^3)}{\rho_{crit}(z)}$, where $\rho_{crit}(z)$ is the critical density of the Universe at redshift $z$}. State--of--the--art N--body hydrodynamical simulations have been proven useful in increasing our theoretical understanding of such effects \cite{ycui}. In this work, we deploy The300-NIKA2 twin samples (TSs) \cite{paliwal} to study the scatter induced on mass ($M$) and integrated SZ Compton parameter ($Y$) due to projection and construct 3D ($Y-M$) scaling laws at different $\Delta$. We also investigate for the optimal overdensity at which the scatter due to the projection effect is minimised. 

\indent This paper is structured as follows: Section \ref{database} gives a brief overview of the observational catalog (NIKA2 SZ Large Program catalog \cite{lpsz}) along with the simulation database ({\sc The Three Hundred}, henceforth The300 \cite{ycui}) and how they are combined to create the so--called TSs. Section \ref{projections} includes the study of the impact of projection on the inference of $M$ and $Y$. Section \ref{sl} describes the construction and analysis of the 3D scaling--laws. Finally, Section \ref{scope} discusses and summarises the inferences drawn from our analyses.
%%%%%%%%%%%%%%%%%%%%%%%%%%%%%%%%%%%%%%%%%%%%%%%%%%%%%%%%%%%%%%

\section{Data}
\label{database}

\subsection{The NIKA2 Sunyaev–Zeldovich Large Program and The300}

The NIKA2 Sunyaev–Zeldovich Large Program (LPSZ) is a part of the NIKA2 (\cite{n1,n2}) guaranteed time. It is essentially a high--resolution ($17.6^{\prime \prime}$ and $11.1^{\prime \prime}$ at  $150 \ \mathrm{GHz}$ and $260 \ \mathrm{GHz}$, respectively) follow--up of 38 SZ--selected galaxy clusters from the Planck \cite{p27} and Atacama Cosmology Telescope (ACT) \cite{hass} catalogs. It spans a high redshift range of $0.5 < z < 0.9$ and an order of magnitude in mass with the cluster masses, $ 3 \leqslant M_{500} / 10^{14} \ \mathrm{M}_{\odot} \leqslant 11$. The combination of prior knowledge of the parameter $Y$ and high--resolution data from NIKA2, enables the precise inference of the pressure profiles of the clusters and consequently their dynamical state and morphology.

To complement the sensitive, high--resolution NIKA2 data with simulations, we utilise The300 project. It consists of simulations of a set of 324 Lagrangian regions of radius $15 \ h^{-1}{\mathrm {Mpc}}$, each centred on a galaxy cluster with $M_{500} > 4.6\times 10^{14} \ h^{-1} {\mathrm M_{\odot}}$. All the regions are simulated at high resolution in $128$ snapshots between redshifts $z=0$ and $z=17$ using three hydrodynamical codes: a standard smooth--particle--hydrodynamic (SPH) code \texttt{GADGET--MUSIC}, and two modern SPH codes \texttt{GADGET--X}, and \texttt{GIZMO-SIMBA}. The technical details about the codes, the initial conditions and box sizes related to these codes can be found in \cite{ycui}. For this work, we use the data products from the \texttt{GADGET--X} simulations. These simulations provide rich, multi--wavelength information about the clusters and their environment. These include mock maps (X--ray, optical, gravitational lensing, radio, and SZ Compton parameter) and 3D Intra Cluster Medium (ICM) profiles (pressure, density, temperature, etc.) of each cluster.
%-------------------------------------------------------------
\subsection{The300--NIKA2 twin samples}

We selected some clusters from The300 simulations that mimic the mass and redshift ranges of the LPSZ clusters, to generate the so--called TSs. At first, we identified snapshots in the simulations whose redshifts matched the median value of the redshift bins of the LPSZ. Then we cross--matched the following properties of the clusters in LPSZ to generate 3 TSs: total mass ($\rm{TS}_{\rm{TM}}$), hydrodynamic mass ($\rm{TS}_{\rm{HSE}}$), and integrated Compton parameter at $r_{500}$ ($\rm{TS}_{\rm{Y}}$). Each of these is a unique representation of the LPSZ in the given criteria. The details of these TSs and their creations can be found in \cite{paliwal}. In this work, we use the simulated Compton parameter ($y$) maps and total project mass maps of the TSs. Both of these maps are available in 29 projections with both, the size and the integration depth of the maps being $4 \times r_{200}$, to study the projection effects. We also use the 3D ICM radial mass profiles, which are simply the sum of the mass of all particles within a sphere as a function of radius, to compare the projected masses with.

%------------------------------------------------------------

\section{Projection effects}
\label{projections}

To infer the scatter induced on mass estimates due to projection, at different apertures, we use the 29 projections of the total projected mass maps. The apertures at which we estimate the integrated quantities are designated using the overdensity described in Section \ref{intro}. $M_{\Delta}$ and $r_{\Delta}$ correspond to the spherical mass and radius obtained from the 3D mass profiles of the simulations and $\rho_{crit}(z)$ is the critical density estimated using the simulation cosmology model (Planck 2015 \cite{pcosmo}).

\begin{figure*}
\centering
\includegraphics[width=0.85\textwidth]{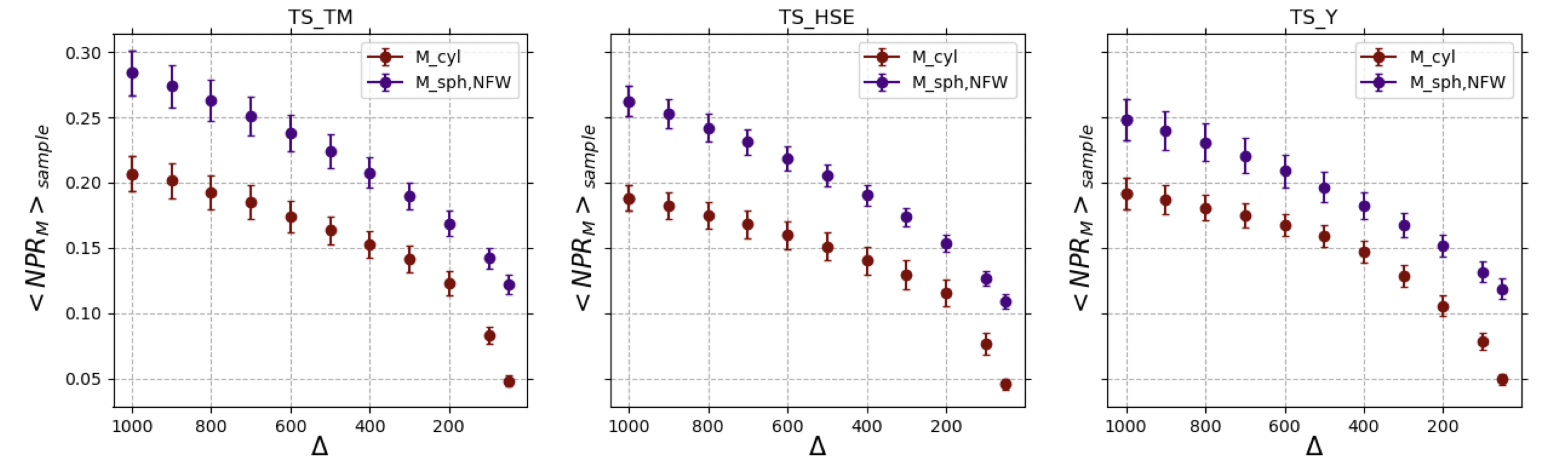}
\includegraphics[width=0.85\textwidth]{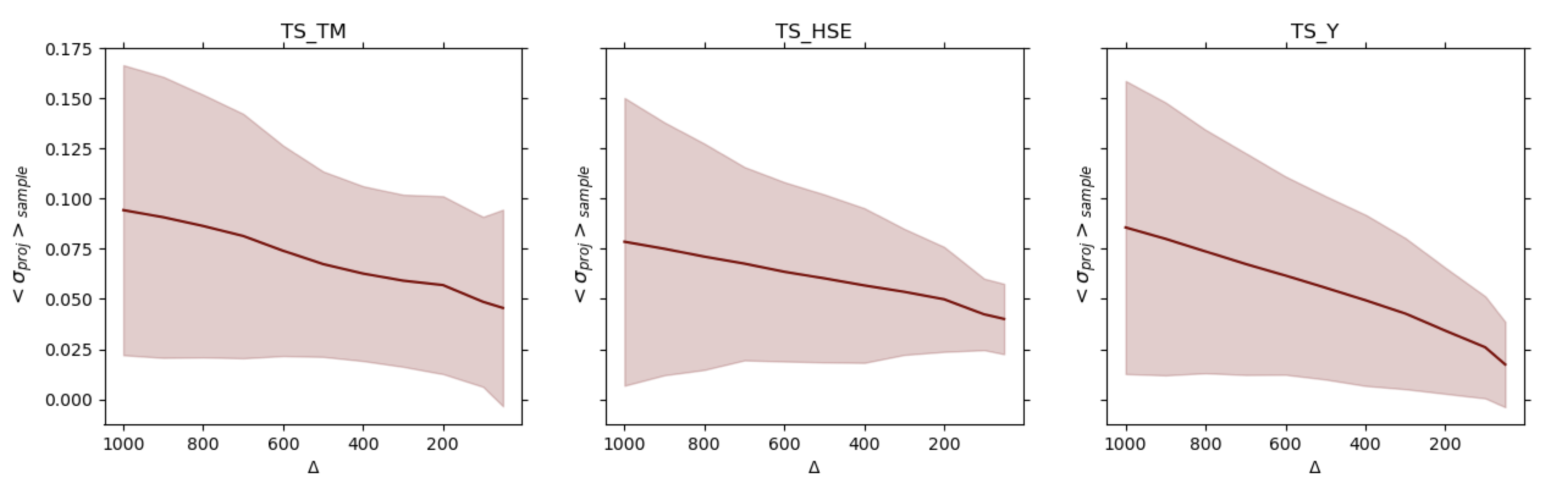}
\caption{Scatter induced on mass (top) and integrated Compton parameter (bottom), for all three TSs. The scatter can be seen to decrease with decreasing overdensity for all samples.}
\label{massscat}
\end{figure*}

The mass is inferred in two ways. First, the total projected mass maps are directly integrated within the apertures defined above to get the total cylindrical mass $M_{\mathrm{cyl}}$. For the second estimate, surface mass density profiles $\Sigma (r)$ are estimated using the mean value, within a given annulus, of the projected mass maps. These profiles, with the dispersion in each aperture as the error, are fitted to the projected Navarro–Frenk–White (NFW) mass profiles \cite{pnfw} using \texttt{COLOSSUS} \cite{col}. The best--fit parameters from these results are used to de--project the profiles to spherical NFW mass profiles \cite{nfw}. From these de--projected mass profiles, spherical masses within designated apertures are obtained and these are referred to as $M_{\mathrm{{sph,NFW}}}$. Both these masses are estimated for all 29 lines of sight. The normalised percentile range ($NPR$) in mass, across the different LoS, is defined as 
    $\rm{NPR} = (\rm{LoS}_{P84} - \rm{LoS}_{P16}) / \rm{LoS}_{median}$ , where P84, P16, and median are the 84th percentile, 16th percentile, and median of the mass distribution along the different LoS. We estimate the $NPR$ for each cluster at different apertures to infer the projection--induced scatter. The sample average of the $NPR$ for each TS, at different apertures can be seen in the top panel of Fig. \ref{massscat}. As expected, the scatter decreases with decreasing overdensity (increasing distance from cluster centre) since the farther out one goes, the more the integrated quantity will look similar irrespective of which direction the cluster is viewed. It is also notable that the scatter induced on cylindrical mass inferred from the projected map is lower than that of the de--projected, spherical masses.

To calibrate a scaling--law between $Y-M$, the estimates of $Y$ are also required. For this, the $y$ maps were used to obtain the integrated Compton parameter $Y_{\mathrm{cyl}}$ by directly integrating the maps within a given aperture. The sample average of the projection--induced scatter on the common logarithmic values of $Y_{\mathrm{cyl}}$ can also be seen in the bottom panel of Fig. \ref{massscat}. As in the case of integrated masses, the scatter here decreases with increasing radial distance from the cluster centre. In order to verify if this scatter in  $Y_{\mathrm{cyl}}$ is correlated with the cluster dynamical state, we test the correlation between the scatter and dynamical state indicator: relaxation parameter $\chi$, estimated in \cite{luca}, at $\Delta=500$. We find no correlation between the two, with correlation coefficients being $-0.01, \ -0.01, \ \mathrm{and}\ -0.02$ for $\rm{TS}_{\rm{TM}}$, $\rm{TS}_{\rm{HSE}}$, and $\rm{TS}_{\rm{Y}}$, respectively. 
To test if the correlation is negligible due to integrating the $y$ signal, we also test the correlation between the scatter in $y$ and $\chi$ and we find a significant correlation, with the correlation coefficients estimated to be $-0.66, \ -0.65, \ \mathrm{and}\ -0.65$ for $\rm{TS}_{\rm{TM}}$, $\rm{TS}_{\rm{HSE}}$, and $\rm{TS}_{\rm{Y}}$, respectively.

%%%%%%%%%%%%%%%%%%%%%%%%%%%%%%%%%%%%%%%%%%%%%%%%%%%%%%%%%%%%%%%%%%%%%
\section{3D scaling laws}
\label{sl}

\begin{figure}
\centering
\includegraphics[width=0.95\textwidth]{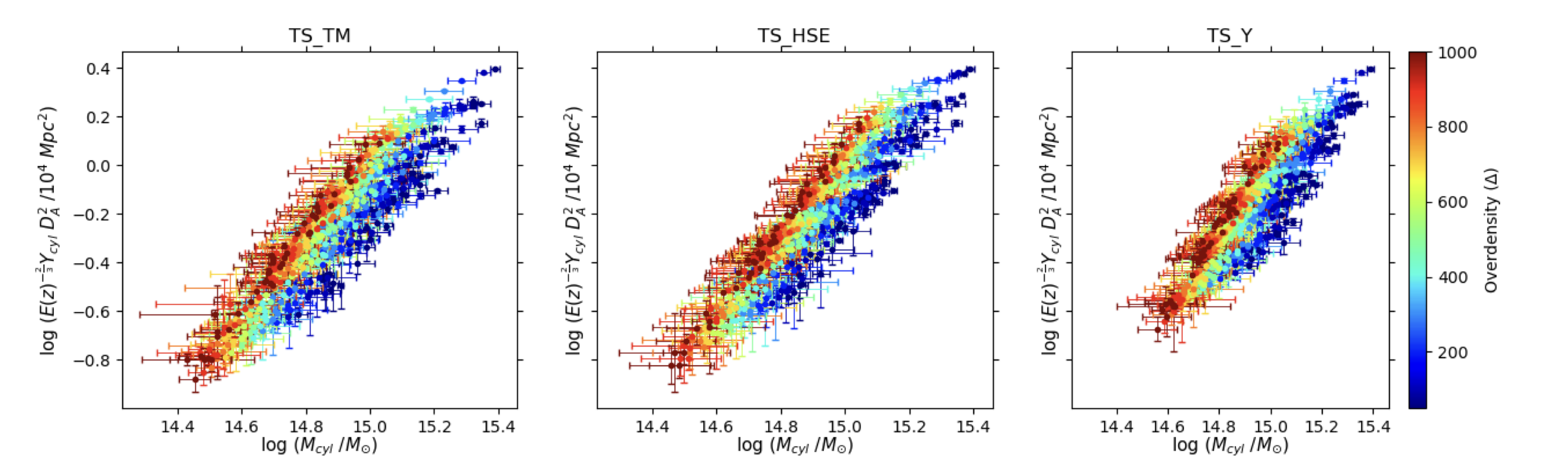} 
\caption{The $Y-M_{\mathrm{cyl}}-\Delta$ distribution for all the TSs. The errors are the projection--induced scatter.}
\label{distri}
\end{figure}

To calibrate a scaling law $Y-M$ for the integrated quantities estimated in Section \ref{projections} we use the power law form described in \cite{psl}

\begin{equation*}
    E(z)^{-\frac{2}{3}} \left[ \frac{D_A^2Y_{\Delta}}{10^{-4} \ \mathrm{Mpc^2}} \right] = A \left[ \frac{M_{\Delta}}{6\times 10^{14} \ \mathrm{M_{\odot}}} \right]^B ;
\end{equation*}
where $E(z)$ and $D_A$ are the Hubble function and angular diameter distance at redshift $z$, respectively. We calibrate two kinds of scaling laws where $M_{\Delta}$ is $M_{\mathrm{cyl}}$ in one case and $M_{\mathrm{{sph,NFW}}}$ in the other and $Y_{\Delta}$ always implies $Y_{\mathrm{cyl}}$. The free parameters of the scaling law are the normalisation $A$, slope $B$, and (common logarithmic) intrinsic scatter $\sigma_{int}$. The intrinsic scatter accounts for the fact that one value of $Y$ never exactly corresponds to one value of $M$ because of the individual traits of a cluster. It must be noted here that typically this scaling law is estimated for $\Delta=500$ but in our case, we have estimated the integrated quantities at varied apertures and therefore we estimate this scaling law at different overdensities. The distribution of these data points for the mass estimate $M_{\mathrm{cyl}}$ can be seen in Fig. \ref{distri} where the distinct evolution of the scaling law with overdensity is evident.

We use \texttt{pylira}\footnote{https://github.com/fkeruzore/pylira}, the python wrapper of the LIRA code \cite{lira} to fit the scaling laws. Since the estimates are purely theoretical, the errors on the integrated quantities are simply the scatter induced due to projection. We assume no redshift evolution in the slope, normalisation, or the intrinsic scatter scatter of the scaling law. Additionally, no prior correlation is assumed in the errors of the integrated quantities.

\begin{figure*}[!ht]
\centering
 \includegraphics[width=0.85\textwidth]{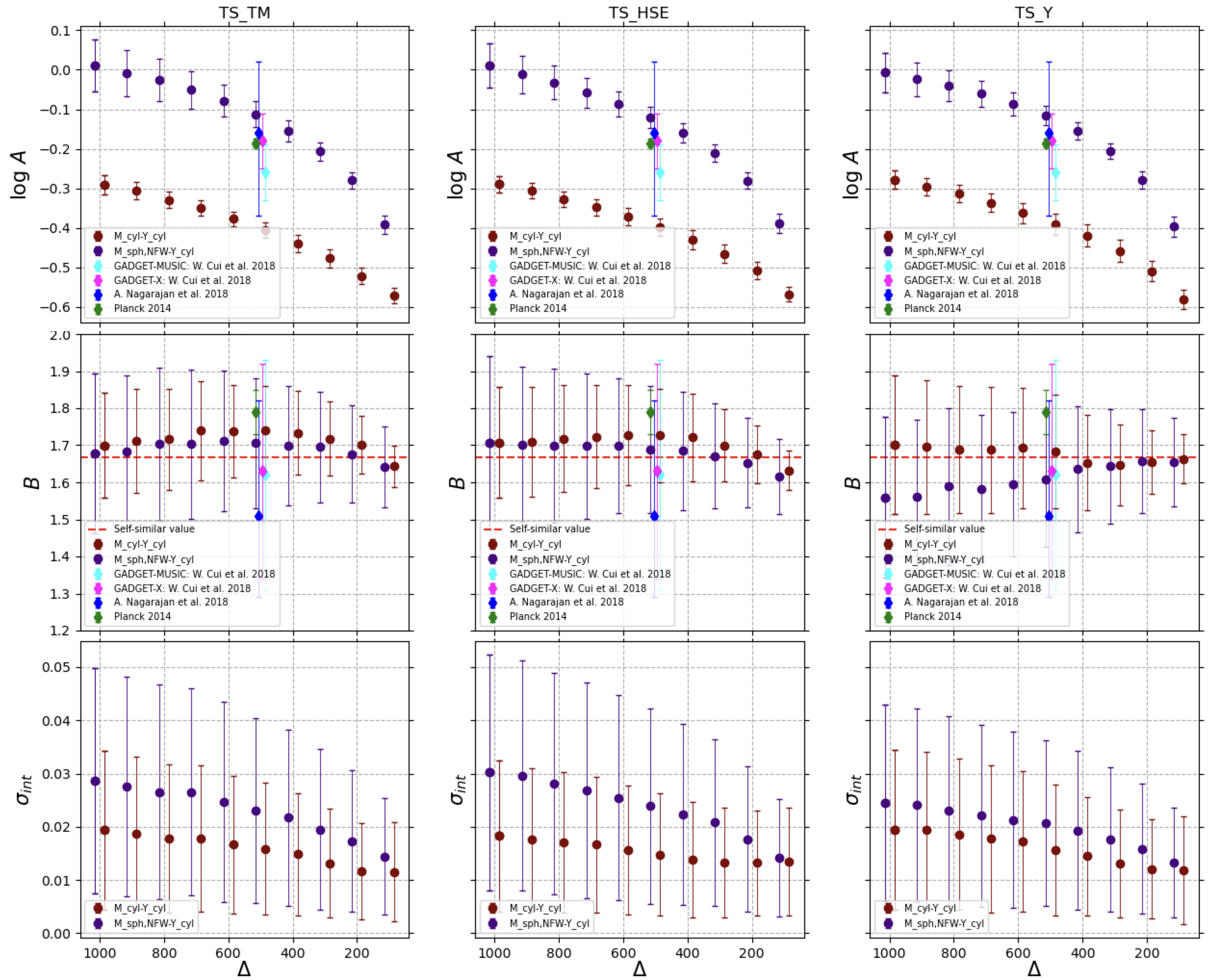}

\caption{The best--fit values of the scaling law parameters at different overdensities are shown for all three TSs in the top (normalization) and central (slope) panels along with the intrinsic scatter in of the scaling law (bottom panel). The errors are the dispersion in the distribution of the parameter estimate.
}
\label{bestfit}   
\end{figure*}

The results for the best--fit values of the scaling law parameters can be seen in Fig. \ref{bestfit}. Given that mass increases with decreasing overdensity and that cylindrical masses are always higher than spherical mass estimates, we see consistently that the normalisation for $M_{\mathrm{cyl}}$ is lower than that of $M_{\mathrm{{sph,NFW}}}$ and that in both cases, it decreases with decreasing overdensity. There is no significant evolution in the slope for either of the scaling laws and the value does not deviate from the self--similarity at any overdensity. Comparing our values with the literature, we find the values of normalisation and slope are in agreement with the \cite{aarti} and the simulation from which the TSs are drawn (\texttt{GADGET--X}). There is a disagreement between our normalisation values and the \texttt{GADGET--MUSIC} simulations but this is expected since these simulations use a different hydrodynamical code. Since the cluster sample used in \cite{psl} is at a lower redshift and different mass range than ours, we find no agreement in our results. Once again the scatter is monotonically decreasing at large radii. Consistent with the scatter induced due to projection, the intrinsic scatter is lower for scaling law estimated using $M_{\mathrm{cyl}}$ compared to that calibrated using $M_{\mathrm{{sph,NFW}}}$. The values of scatter induced due to projection and the intrinsic scatter, are compatible within their error bars.

%%%%%%%%%%%%%%%%%%%%%%%%%%%%%%%%%%%%%%%%%%%%%%%%%%%%%%%%%%%%%%
\section{Discussions and conclusion}
\label{scope}
In this work we have started to investigate, using The300-NIKA2 twin samples, the impact of projection on integrated quantities $Y$ and $M$, at different overdensities. Additionally, we have built and analysed 3D scaling laws. We find no optimal overdensity at which the scatter induced on $Y$ and $M$ due to projection and the intrinsic scatter of the $Y-M$ scaling law is minimised, \textit{i.e.} the scatter always decreases with decreasing overdensity. There is clear evidence that projection--induced scatter on $M$, as well as the $Y-M$ scaling law intrinsic scatter, are lower when using the mass estimate $M_{\mathrm{cyl}}$ as compared to when using $M_{\mathrm{{sph,NFW}}}$. We also find that the projection--induced scatter on $Y$ and the intrinsic scatter of the $Y-M$ scaling law are compatible within their error bars and hence, the former can not be ignored. In conclusion, this work establishes that it is best to observe the cluster as far out as possible in order to minimise the projection--induced scatter on $Y$ and $M$ as well as the intrinsic scatter of the $Y-M$ scaling law and that the two scatters are comparable in value. However, these results come with the caveat that no observational error or instrumental effect has been taken into account and the errors are related purely to projection effects.

\end{document}